\documentclass[aps,prl,10pt,twocolumn,superscriptaddress]{revtex4-2}

\usepackage{spinstab}

\begin{document}

\title{Autonomous feedback stabilization of a cavity-coupled spin oscillator}

\date{\today}

\author{Julian Wolf}
\email{julian.wolf@berkeley.edu}
\author{Olive H. Eilbott}
\author{Joshua A. Isaacs}
\altaffiliation[Present address: ]{Eikon Therapeutics, Hayward, CA 94545, USA}
\affiliation{Department of Physics, University of California, Berkeley, California 94720, USA}
\affiliation{Challenge Institute for Quantum Computation, University of California, Berkeley, California 94720, USA}
\author{Kevin P. Mours}
\altaffiliation[Present address: ]{Max-Planck-Institut f\"ur Quantenoptik, Garching, Germany}
\altaffiliation{Munich Center for Quantum Science and Technology (MCQST), 80799 Munich, Germany}
\affiliation{Department of Physics, University of California, Berkeley, California 94720, USA}
\affiliation{Challenge Institute for Quantum Computation, University of California, Berkeley, California 94720, USA}
\affiliation{Technische Universit\"at Kaiserslautern, 67663 Kaiserslautern, Germany}
\author{Jonathan Kohler}
\altaffiliation[Present address: ]{Vector Atomic, Inc., Pleasanton, CA 94586, USA}
\affiliation{Department of Physics, University of California, Berkeley, California 94720, USA}
\author{Dan M. Stamper-Kurn}
\affiliation{Department of Physics, University of California, Berkeley, California 94720, USA}
\affiliation{Challenge Institute for Quantum Computation, University of California, Berkeley, California 94720, USA}
\affiliation{Materials Sciences Division, Lawrence Berkeley National Laboratory, Berkeley, California 94720, USA}

\begin{abstract}
    We report out-of-equilibrium stabilization of the collective spin of an atomic ensemble through autonomous feedback by a driven optical cavity.
    For a magnetic field applied at an angle to the cavity axis, dispersive coupling to the cavity provides sensitivity to a combination of the longitudinal and transverse spin.
    Coherent backaction by cavity light onto the atoms, conditioned by the optical cavity susceptibility, stabilizes the collective spin state at an arbitrary energy.
    The set point tracking and closed-loop gain spectrum of the feedback system are characterized and found to agree closely with analytic predictions.
\end{abstract}

\maketitle


As in the case of classical systems, the state and evolution of quantum systems can be tailored by feedback control~\cite{Lloyd2000,Mabuchi2005,Zhang2017}.
At a scientific level, the development of a quantum control theory, one that integrates entanglement and non-classical effects of dissipation and measurement, opens a new line of inquiry into non-equilibrium and open quantum systems.
At an applied level, feedback control allows quantum devices to operate robustly, mitigating errors in system preparation and calibration as well as decoherence.
Feedback control underpins important tasks such as error correction in quantum computation~\cite{Shor1995,Steane1996,Geerlings2013} and sensing~\cite{Kessler2014,Dur2014,Zhou2020}, entanglement purification~\cite{Pan2001}, and adaptive measurement~\cite{Armen2002}.

Quantum feedback control can be divided broadly into the two categories of measurement-based and autonomous feedback.
In the measurement-based approach, properties of a quantum system are read out on a classical sensor, the measurement record of which is used by an extrinsic classical control device to alter the ensuing coherence, dissipation, and measurement operations on the system~\cite{Sayrin2011,Vijay2012,Riste2012,CampagneIbarcq2013,Doherty2000}.
The feedback system's design must account for noise and backaction that are intrinsic to quantum measurement.

By comparison, in autonomous (or coherent) quantum feedback the corrective response that steers a quantum subsystem is built into the quantum system itself.
Control is achieved by structuring the drive and dissipation of an open quantum system so that entropy is reliably extracted as the quantum system is steered to the desired final state. 
Examples of such schemes include autonomous error correction in bosonic code spaces~\cite{Gertler2021,Puri2020,Ma2021}, quantum state preparation~\cite{Murch2012,Riste2013,Shankar2013,Lin2013,Leghtas2013,Andersen2016}, optical noise cancellation~\cite{Mabuchi2008}, and generating spin squeezing of atoms in a driven optical resonator~\cite{SchleierSmith2010}.

In this work, we develop a coherent feedback scheme to stabilize the energy of an ensemble of quantum spins.
Our scheme employs optical backaction in a driven cavity to realize closed-loop autonomous feedback.
Under negative-feedback conditions, we observe that cavity spin optodynamics~\cite{Brahms2010,Kohler2017} deterministically steer the collective spin toward a steady-state energy that is set by the frequency of the driving optical field.
By examining both the light that drives the system and the atomic spins that respond to this drive, we quantify the tuning sensitivity as well as the closed-loop gain spectrum of the autonomous feedback system and find close agreement with a theoretical model.


The feedback system comprises the collective spin of an ultracold atomic gas interacting with an optical cavity mode.
In particular, an ensemble of $\Na \approx 1400$ non-degenerate \Rb{} atoms, cooled to around $\SI{3}{\micro\kelvin}$, is trapped predominantly in a single antinode of a standing-wave optical dipole trap (ODT, wavelength $\SI{842}{\nano\meter}$) resonant with a TEM\textsubscript{$00$} mode of an optical Fabry--P\'erot cavity~\cite{Purdy2010, Kohler2017, Kohler2018, Zeiher2021}.
The atoms are initially prepared in the $\ket{{f = 2},{m_f = 2}}$ hyperfine level of the electronic ground state.

\begin{figure}[!tb]
    \begin{center}
    \includegraphics[scale=1]{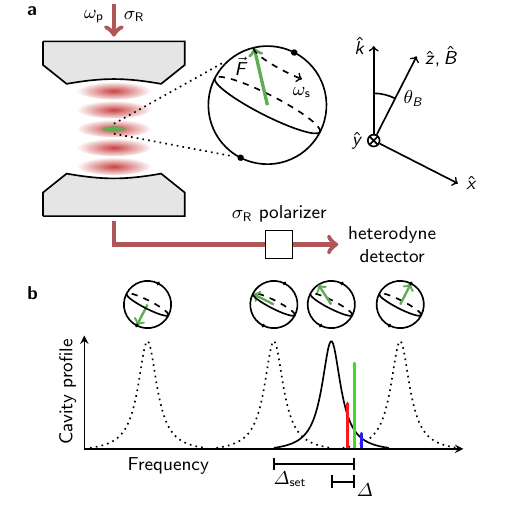}
    \end{center}

    \caption{ \label{fig:schematic}
        Experimental setup.
        \textbf{a.}~%
            Atoms are localized at an antinode of the intracavity pump field, such that they are coupled symmetrically to the cavity mode.
            A magnetic field $\vec B = B \,\hat z$ is applied at an angle $\thetaB$ relative to the cavity axis $\hat k$.
            The cavity is pumped with circularly polarized ($\sigmaR$) light.
            The $\sigmaR$ light emitted by the cavity is sent to a balanced heterodyne detector.
        \textbf{b.}~%
            The projection of the collective spin onto the cavity axis, $\langle \Fk \rangle$, results in a dispersive shift to the cavity resonance.
            For $\thetaB < \SI{90}{\degree}$, positive-energy (negative-energy) spin states shift the cavity resonance, on average, by $\DeltaFast > 0$ ($\DeltaFast < 0$).
            For pump light detuned from the cavity resonance (vectical green line), the Stokes and anti-Stokes sidebands (red and blue lines) are filtered by the cavity, and have different amplitudes, resulting in net energy transfer between the light and the collective spin.
    }

\end{figure}

The atomic ensemble interacts strongly with a second TEM\textsubscript{$00$} cavity mode (the ``pump'' mode) whose frequency $\omegaC$ is detuned from the atomic \Dtwo{} transition (frequency $\omegaA$, wavelength $\SI{780}{\nano\meter}$) by $\DeltaCA \equiv \omegaC - \omegaA = 2 \pi \times \SI{-35}{\giga\hertz}$.
The half-linewidth of the cavity at the \Dtwo{} wavelength is $\kappa / 2 \pi = \SI{1.82}{\mega\hertz}$.
We minimize the effects of the spatial dependence of the atom--cavity coupling by trapping atoms at a location where the trapping field and pump field antinodes coincide (\fig{schematic}a)~\cite{Purdy2010}.
The symmetric coupling of the atoms to the cavity field allows the ensemble to be addressed in terms of a total (dimensionless) spin $F = 2 \Na \approx 2800$ and a mean dispersive cavity--atom coupling $\gc = g_0^2 / \DeltaCA$, where $g_0 = 2 \pi \times \SI{13}{\mega\hertz}$ is the vacuum Rabi coupling of a single atom to the pump mode, averaged over the atom's motion in the ODT~\cite{Purdy2010}.

For a cavity pumped with circularly polarized ($\sigmaR$, or $\sigma_-$ for an external field pointing along the cavity axis) light, the dynamics of the system are governed by the Hamiltonian~\cite{Brahms2010} (Appendix~A)
\begin{equation} \label{eqn:hamiltonian-general}
    \hat H =
        - \hbar \DeltaPC \hat c^\dag \hat c
        + \hbar \omegaS \Fz
        + \hbar \gc \big[ \alpha_0 \Na - \alpha_1 \Fk \big] \,\hat c^\dag \hat c,
\end{equation}
written in a frame rotating at the pump frequency $\omegaP$.
Here, $\DeltaPC \equiv \omegaP - \omegaC$ is the pump--cavity detuning, $\hat c$ is the cavity pump mode annihilation operator, and $\omegaS = g_F \muB B / \hbar = \SI{300}{\kilo\hertz}$ is the Larmor frequency (where $g_F$ is the Land\'e $g$-factor, $B$ the applied magnetic field strength, and $\muB$ is the Bohr magneton).
The spin projections $\Fz$ and $\Fk$ in \eqn{hamiltonian-general} are defined by the applied magnetic field $\vec B = B \hat z$ and the cavity axis $\hat k$ (\fig{schematic}a).
The constants $\alpha_0 = \frac 2 3$ and $\alpha_1 = \frac 1 6$ describe the scalar and vector interactions, respectively, between the atoms and the cavity pump field (Appendix~A).
We have treated atom--cavity interactions as purely dispersive, accounting for $|\DeltaCA|$ being large compared to the atomic half-linewidth ($\gamma = 2 \pi \times \SI{3}{\mega\hertz}$) and to the collective atom--cavity coupling strength $\sqrt{N} g_0$.

The collective atom--cavity interaction is the sum of two terms.
A scalar (spin-independent) dispersive atom--light interaction shifts the cavity resonance frequency proportional to atom number $\Na$.
With $\Na$ being constant during the few-$\si{\milli\second}$ duration of the spin feedback experiments, it is useful to absorb this static frequency shift into an effective constant pump--cavity detuning $\DeltaSlow \equiv \DeltaPC - \gc \alpha_0 \Na$.
In addition, a vector (spin-dependent) atom--cavity interaction shifts the resonance frequency of the $\sigma_-$ cavity mode by an amount proportional to the spin component $\Fk$ (Appendix~A).

We now outline how this quantum system autonomously includes the essential elements of a feedback control system.
In such a control system, a control variable, which represents the state of the plant (the subsystem to be controlled), is measured by a sensor.
A comparator generates an error signal as the difference of the sensor output and an externally determined set point.
A controller conditions the error signal and acts on the plant.
Under proper negative-feedback conditions, the control variable is stabilized unconditionally.

Identifying $\Fk = \Fx \sin \thetaB + \Fz \cos \thetaB$, with $\thetaB$ being the angle between the applied magnetic field and $\hat k$ (\fig{schematic}a), we observe that the cavity is sensitive both to the longitudinal spin $\Fz$ (equivalently, the bare spin energy) and the transverse spin $\Fx$.
The longitudinal spin $\Fz$ plays the role of the control variable, and the cavity shift proportional to $\Fz$ acts as a coherent sensor.
Accounting for this shift, the net detuning of the pump light from the cavity, averaged over the fast effects of Larmor precession, is given by $\hat \DeltaFast \equiv \DeltaSlow + \gs \Fz \cos \thetaB$, with $\gs \equiv \alpha_1 \gc$.
The system Hamiltonian \eqn{hamiltonian-general} can now be rewritten as
\begin{equation} \label{eqn:hamiltonian-feedback}
    \hat H =
        - \hbar \hat \DeltaFast \hat c^\dag \hat c
        + \hbar \omegaS \Fz
        - \hbar \gs \hat c^\dag \hat c \,\Fx \sin \thetaB.
\end{equation}
The net detuning $\hat \DeltaFast$ represents the control system error signal, proportional to the difference between the instantaneous value of $\Fz$ and the externally determined longitudinal spin set point
\begin{equation} \label{eqn:Fz-final}
    \Fset
        = -\frac{\DeltaSlow}{\gs \cos \thetaB}.
\end{equation}

The final term in \eqn{hamiltonian-feedback} completes the autonomous control system, serving as the feedback controller.
As described in Refs.~\cite{Kippenberg2008,Brahms2010,Kohler2018}, the Larmor precessing transverse spin modulates the cavity field intensity through the spin-dependent dispersive interaction. 
In turn, this modulation, conditioned by cavity dynamics, acts resonantly on the precessing spin and alters its energy.

As in the case of cavity optomechanics~\cite{Marquardt2007}, the resulting energy dynamics of the spin ensemble can be described in terms of cavity-induced sideband asymmetry~\cite{Vuletic2000}.
Modulation of the cavity resonance by the precessing spin shifts optical power from the pump light into first-order frequency sidebands with optical frequencies $\omegaP \pm \omegaS$.
While a free-space modulator would generate sidebands with equal power, here, the cavity spectrum induces a sideband asymmetry (\fig{schematic}b).
For a pump blue-detuned from cavity resonance ($\DeltaFast > 0$), the cavity induces stronger emission on the Stokes (red) sideband, reducing the net energy of the optical pump and, in turn, increasing the energy of the spin ensemble.
Similarly, a pump red-detuned from cavity resonance ($\DeltaFast < 0$) reduces the energy of the spin ensemble.
With the correct sign of $\gs \cos \thetaB$ in \eqn{Fz-final}, the response of the spin ensemble in either case brings $\DeltaFast$ closer to zero.
The system arrives at a stable steady state, with $\langle \Fz \rangle = \Fset$, that is determined by the pump frequency $\omegaP$ (through $\DeltaSlow$) and that, notably, is independent of the initial state of the collective spin and insensitive to many perturbations.

\begin{figure}[!bt]
    \begin{center}
    \includegraphics[scale=1]{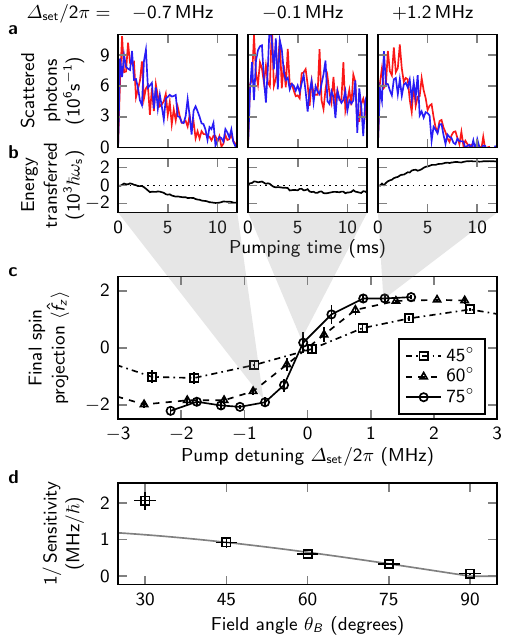}
    \end{center}

    \caption{ \label{fig:dc-pulling}
        Response of the cavity--spin system to a constant pump tone.
        \textbf{a.}~%
            Power asymmetry between the Stokes (red) and anti-Stokes (blue) sidebands of the optical cavity emission records the exchange of energy between the cavity field and the spin ensemble.
            Data are recorded at $\thetaB = \SI{75}{\degree}$.
        \textbf{b.}~%
            Integrating this asymmetry yields the net energy transfer $\delta E(t)$ to the spin ensemble vs.\ time.
            For $\DeltaSlow < 0$ (left panel), light is preferentially scattered into the anti-Stokes sideband, pumping energy out of the spin mode; for $\DeltaSlow > 0$ (right panel), energy is pumped into the spin mode.
            For $\DeltaSlow \approx 0$, light is scattered equally into the two sidebands, yielding no net energy transfer.
        \textbf{c.}~%
            The final longitudinal spin after $\SI{2}{\milli\second}$ of feedback is controlled by $\DeltaSlow$.
            For large $\sin \thetaB$ (\emph{e.g.}, $\thetaB = \SI{75}{\degree}$, circles; and $\thetaB = \SI{60}{\degree}$, triangles), the final spin tunes over its entire range, up to $\langle f_z \rangle = \pm 2$ for sufficiently large $\DeltaSlow$.
            For smaller $\sin \thetaB$ (\emph{e.g.}, $\thetaB = \SI{45}{\degree}$, squares), the collective spin dephases before it can be pulled all the way to either extreme.
            Error bars represent standard errors on the mean, averaged over $7$ repetitions of the experiment.
        \textbf{d.}~%
            The system's sensitivity to pump detuning, $2 \pi \dd \,\langle \hat f_z \rangle / \dd \DeltaSlow |_{\DeltaSlow = 0}$ (black squares: measurement; gray line: \eqn{Fz-final} prediction), depends on magnetic field orientation $\thetaB$.
            Error bars represent fit uncertainties.
    }

\end{figure}

We first confirm experimentally that the spin ensemble is autonomously stabilized to a state determined by the external set point.
To this end, the collective spin is initiated to $\langle \Fz(t = 0) \rangle = 0$ using a coherent rf $\pi/2$-pulse at drive frequency $\omegaS$, such that $\DeltaFast(t = 0) = \DeltaSlow$.
The cavity is then pumped with light at a constant $\DeltaSlow$ and allowed to evolve.
The light emitted by the cavity is measured on a balanced heterodyne detector~\cite{Brooks2012,Brahms2012,Zeiher2021} (total detection efficiency $\edet = \SI{2.2}{\percent}$), allowing the power in the Stokes and anti-Stokes sidebands to be detected as independent time traces (\fig{dc-pulling}a).
The difference in the power of the two sidebands directly measures the instantaneous energy transfer from the pump light to the collective spin.
The cumulative sum of this difference measures the total energy $\delta E(t)$ added to the collective spin, leading up to time $t$.
As shown in \fig{dc-pulling}a and~b, an initial $\DeltaSlow < 0$ leads to an enhancement of the anti-Stokes sideband and a net energy transfer $\delta E < 0$, driving the spin to a low energy state, while $\DeltaSlow > 0$ has the opposite effect.

The spin state achieved after long evolution times under autonomous feedback, at a given $\DeltaSlow$, is shown in \fig{dc-pulling}c.
Here, we measure the longitudinal spin by terminating the feedback, reorienting the magnetic field along $\hat k$, and measuring the spin-dependent cavity shift (Appendix~B).
For each $\thetaB$, the measured response shows a tuning range, centered about $\DeltaSlow = 0$, within which the steady-state spin energy varies linearly with $\DeltaSlow$.
The sideband-based energy transfer measurements show the same trend as the spin measurements, through the relation $\hbar \omegaS \langle \Fz \rangle(t) = \delta E(t)$, but are found to be less precise.
By fitting the response curves to an analytical model (discussed below and in Appendix~C), we determine the linear sensitivity of the steady-state spin to $\DeltaSlow$ near $\DeltaSlow = 0$.
This linear sensitivity (\fig{dc-pulling}d) matches well to the prediction of the set point equation (\eqn{Fz-final}) for a range of field angles.

Outside the linear tuning range $|\DeltaSlow| > |2 \Na \gs \cos\thetaB|$, one would expect the feedback system to rail, driving the spin ensemble to one of its extremal energy states.
Such railing is observed for $\thetaB \gtrsim \SI{55}{\degree}$.
Here, the cavity field modulation amplitude, proportional to $\sin \thetaB$, is large, driving the spin quickly to its steady state.
In contrast, for shallower angles ($\thetaB \lesssim \SI{55}{\degree}$), the collective spin undergoes dephasing during feedback, reducing its total magnitude to $|| \vec F || < 2 \Na$ before the system can reach its steady state.


Next, we investigate the dynamical response of the autonomous feedback system.
Considering the Hamiltonian of \eqn{hamiltonian-feedback} and adding terms accounting for pumping into and leakage out of the cavity mode, the cavity field evolves according to
\begin{equation} \label{eqn:cavity-field}
    \frac{\dd}{\dd t} \hat c
        = \ii \big( \DeltaSlow + \gs \Fk \big) \hat c
            - \kappa \hat c
            + \kappa \eta,
\end{equation}
where $\eta$ is the coherent-state amplitude of the field pumping the cavity.
The energy of the collective spin, meanwhile, evolves according to
\begin{equation} \label{eqn:F-pulling}
    \frac{\dd}{\dd t} \Fz =
        \gs \hat c^\dag \hat c \,\Fy \,\sin \thetaB.
\end{equation}
For $\omegaS \gg \gs$, as in our experiment, the optodynamical Larmor frequency shift (analogue of the optomechanical spring shift)~\cite{Brahms2010} is small and the transverse spin can be approximated as $\Fy = \Fperp \sin \omegaS t$; in practice, this approximation relies on terms with nontrivial commutation relations only entering in at a higher order than is being considered, and amounts to treating $\Fy$ as equal to its expectation value.

We expect a cavity field comprising a carrier at frequency $\omegaP$ and sidebands at frequencies $\omegaP \pm \omegaS$: $\hat c = \hat c_0 + \hat c_+ e^{-\ii \omegaS t} + \hat c_- e^{\ii \omegaS t}$.
In the limit of small modulation depth $(\gs / 2 \kappa) F_\perp |\sin \thetaB| \ll 1$, the amplitudes and phases of the sidebands $\hat c_\pm$ can be calculated directly from \eqn{cavity-field}.
Inserting this solution for the cavity field into \eqn{F-pulling}, we find that the cavity resonance frequency is pulled toward its set point at a damping rate given as
\begin{equation} \label{eqn:damping-rate}
    \dampingRate \equiv
            \frac{1}{\DeltaFast} \frac{\dd \DeltaFast}{\dd t}
        =   -2 \Fperp^2 \sin^2 \thetaB \cos \thetaB
            \frac{\gs^3}{\kappa^3} \omegaS \nbar.
\end{equation}
Here, $\nbar \equiv \nmax \,\kappa^2 / (\kappa^2 + \DeltaFast^2)$ with $\nmax = |\eta|^2$.
This simple model allows for straightforward simulation of how the system will act under a variety of conditions.


\begin{figure}[!th]
    \begin{center}
    \includegraphics[scale=1]{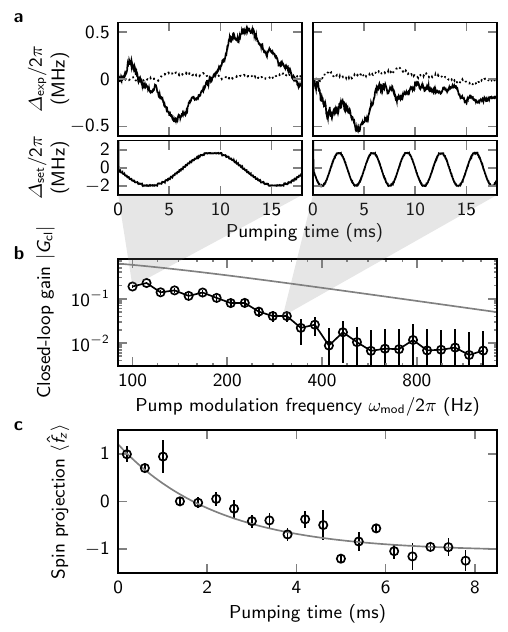}
    \end{center}

    \caption{ \label{fig:ac-response}
        Response of the autonomous cavity--spin feedback system to impulses.
        \textbf{a.}~%
            The real-time cavity shift $\DeltaExp(t)$ due to the collective spin reveals that the system responds directly to a time-dependent effective detuning $\DeltaSlow(t)$.
            Dotted lines show the response of the cavity with no atoms present, as a baseline for measurement noise.
        \textbf{b.}~%
            The measured closed-loop gain of the system is measured as determined by \eqn{modulation-response} (black circles).
            Error bars derive primarily from detection shot noise.
            The expected gain for a perfect pure-integrator system (\eqn{transfer-function}) is also shown (gray line).
        \textbf{c.}~%
            In a different set of experiments, the collective spin is prepared near $\langle \hat f_z \rangle = +1$ and the pump is left on at a constant detuning $\DeltaPC$ for a varying amount of time, after which the spin state is measured nondestructively (black circles).
            This allows the spin's trajectory as the system approaches its equilibrium condition near the $\langle \hat f_z \rangle = -1$ to be reconstructed.
            An exponential fit is used to extract the closed-loop damping rate $\dampingRate = \SI{450 \pm 60}{\per\second}$ (gray line).
            Error bars represent standard errors on the mean, taken over $20$ repetitions of the experiment.
        All data are recorded at $\thetaB = \SI{60}{\degree}$.
    }

\end{figure}

We probe the dynamics of our feedback system in two experiments.
First, we characterize the system's closed-loop transfer function by pumping the cavity with a time-varying tone $\DeltaSlow(t) = \DeltaSlowAmp \sin \omegaMod t$ and measuring the response $\delta E(t)$.
We predict the shift to the cavity resonance due to the spin response to be $\DeltaExp(t) = -\gs \cos \thetaB \delta E(t) / \hbar \omegaS$; this can be compared to the setpoint detuning $\DeltaSlow(t)$  (\fig{ac-response}a).
At each modulation frequency $\omegaMod$, the closed-loop gain $\gain$ is calculated as the ratio between the response and the perturbation (\fig{ac-response}b, black circles):
\begin{equation} \label{eqn:modulation-response}
    \gain[\omegaMod] =
        \frac{2 \ii}{\DeltaSlowAmp T} \int_0^T \dd t \,\DeltaExp(t) \,\exp(-\ii \omegaMod t),
\end{equation}
where $T = 2 \pi s / \omegaMod$, for integer $s$.
For a pure integrator system such as ours with damping rate $\dampingRate$, we expect a closed-loop gain of
\begin{equation} \label{eqn:transfer-function}
    \gain[\omega] =
        \frac{\ii (\dampingRate / \omega)}{1 + \ii (\dampingRate / \omega)},
\end{equation}
which should describe the system well for $\DeltaSlowAmp \ll \kappa, |2 \Na \gs \cos \thetaB|$ (\fig{ac-response}b, gray line).
Our measurements qualitatively match the amplitude response predicted by \eqn{transfer-function}, but the data quality is limited by a signal-to-noise ratio of approximately $0.1$\,--\,$0.3$ (dominated by shot noise on the detection of the sideband fields, which is exacerbated by the low $\edet$~\cite{SNR}) as well as by the large $\DeltaSlowAmp$ used for this measurement (which leads to the collective spin being pulled away from the linear feedback regime $\Fperp \approx 2 \Na$; see below).
These limitations prevent us from measuring the phase response of $\gain$, and from making rigorous comparisons to theory.

Second, and more quantitatively, we characterize the impulse response function of the feedback system.
Here, we initialize the collective spin near $\langle \hat f_z \rangle = +1$, and then suddenly turn on the pump light in order to impose feedback with a set point of $\Fset / \Na = -1$ (\fig{ac-response}c).
Time-resolved direct spin measurements track the system evolution toward the set point (Appendix~B).
For regions over which $\dampingRate$ is approximately constant (namely, $| \langle \hat f_z \rangle | \leq 1$), \eqn{damping-rate} states that $\langle \Fz \rangle$ should approach $\Fset$ exponentially.
Fitting the observed spin response to an exponential relaxation yields an empirical value of $\dampingRate = \SI{450 \pm 60}{\per\second}$ (\fig{ac-response}c).
For the same experimental parameters ($\thetaB = \SI{60}{\degree}$, $\omegaS = 2 \pi \times \SI{300}{\kilo\hertz}$, $\nmax = 2.4$, $\Na = 1100$), \eqn{damping-rate} predicts $\dampingRate = \SI{1600}{\per\second}$.
The disagreement between our measurement and this simple theory prediction is due, in part, to the large modulation depth imposed by the precessing spins in our experimental setting: here, $(\gs / 2 \kappa) \Fperp \sin \thetaB = 0.4$, which warrants the inclusion of terms corresponding to higher-order scattering events (measured as sidebands at $\pm 2 \omegaS$).
Accounting for these corrections reduces the expected gain to $\dampingRate = \SI{790}{\per\second}$ (Appendix~C).
The analytical model also does not account for the dephasing of the spin ensemble.
Constructing an accurate model for dephasing in our system is not straightforward, but any form of dephasing will have the effect of decreasing $\Fperp$, and thus $\dampingRate$, which may explain the remaining discrepancy.


In this work, we have shown that autonomous feedback generated by optical backaction of a driven cavity onto a spin ensemble stabilizes the ensemble energy at an energy determined by the cavity pump frequency.
The optical cavity emission provides a real-time record of the feedback dynamics.
In future work, information from this real-time optical signal may also be used to enhance the feedback stabilization through additional measurement-based feedback~\cite{Hofman1998,Wang2001}.

This system can equivalently be described as autonomous feedback stabilization of the optical cavity's resonance frequency.
From this viewpoint, the control variable is $\DeltaFast$.
The spin ensemble now plays the part of the controller by which the cavity is autonomously tuned to be in resonance with the light with which it is driven.

Our feedback setup stabilizes the spin ensemble to a specific value of the longitudinal spin, but does not control the phase at which this spin undergoes Larmor precession because of the time translation symmetry of our scheme.
Future work may investigate methods for more complete control of the quantum spin state, \emph{e.g.}, applying phase coherent modulation at the Larmor frequency, either to the optical pump field or to an applied magnetic field, so as to stabilize the Larmor precession phase.

Another target for future investigation is the fluctuation of the spin ensemble under steady-state feedback.
In steady state, the ensemble should respond to the quantum noise of the cavity field, generating fluctuations in the longitudinal spin as well as the Larmor precession phase.
At the same time, coherent feedback suppresses longitudinal spin fluctuations.
The balance between quantum-optical fluctuations and coherent dissipation, achieved in the steady state and away from thermal equilibrium, may be revealed in the spectrum of the cavity output.
However, in our current setup, technical noise on $\omegaC$, the pump light spectrum, and optical detectors obscures this quantum noise signature.


\begin{acknowledgments}

We acknowledge support from the National Science Foundation Quantum Leap Challenge Institutes program (Grant No.~OMA-2016245), from the National Science Foundation (Grant No.~PHY-1707756), from the Air Force Office of Scientific Research (Grant No.~FA9550-19-1-0328), and from Army Research Office through the Multidisciplinary University Research Initiative program (Grant No.~W911NF-20-1-0136).
The contributions of J. A. I. are funded by the Heising-Simons Foundation (Grant No.~2020-2479).
The contributions of O. H. E. are supported by the National Science Foundation Graduate Research Fellowship Program (Grant No.~DGE-175281).

\end{acknowledgments}

\bibliographystyle{unsrt}
\bibliography{references}

\begin{thebibliography}{10}
\bibitem{SNR}
The signal being measured looks like the difference in detected photons between
  the Stokes and anti-Stokes sidebands, which, over the course of a
  $\tau_\text{meas} = \SI{20}{\milli\second}$ experiment, is in the range of
  $10$\,--\,$40$. The shot noise at the heterodyne detector, which is the
  dominant noise source, is calculated as $\sqrt{2 \kappa \nbar
  \tau_\text{meas} \edet} = 140$ photons.
\end{thebibliography}

\newpage
\appendix

\section{A. Derivation of the autonomous spin stabilization Hamiltonian}
\label{sec:appendix:hamiltonian}

The Hamiltonian for the autonomous spin stabilization system can be written, generically, as a sum of cavity, spin, and interaction terms: $\hat H = \Hcav + \Hspin + \Hint$.

In the lab frame, the cavity Hamiltonian is simply given by $\Hcav^\text{lab} = \hbar \omegaC \hat c^\dag \hat c$.
We find it helpful to move to a frame rotating at the frequency $\omegaP$ of the cavity pump laser, such that
\begin{equation} \label{eqn:Hcav}
    \Hcav
        = -\hbar \DeltaPC \hat c^\dag \hat c
        = -\hbar \DeltaPC \hat n,
\end{equation}
where $\DeltaPC \equiv \omegaP - \omegaC$ is the pump--cavity detuning.
Here, the cavity annihilation operator $\hat c$ and the photon occupation operator $\hat n$ include light of both right- and left-handed circular polarizations.
Although the left- and right-handed cavity modes interact differently with the atomic ensemble, their bare energies are approximately degenerate, and here they can be considered together as $\hat n = \hat n_\text{R} + \hat n_\text{L}$.

For an ensemble of noninteracting atoms indexed $i$ at positions $\vec r_i$ and spin projection $f_z^{(i)}$ along the direction of the magnetic field, the spin Hamiltonian is given by $\Hspin = \sum_i \hbar \omegaS(\vec r_i) \,\hat f_z^{(i)}$.
Here, the local spin precession frequency is given by $\hbar \omegaS(\vec r) = g_F \muB |\vec B(\vec r)|$, where $g_F$ is the Land\'e $g$-factor and $\muB$ is the Bohr magneton.
For a localized ensemble of atoms, the magnetic field is approximately constant, such that this can be rewritten in terms of an average spin precession frequency $\omegaS$ and a total spin projection $\Fz = \sum_i \hat f_z^{(i)}$:
\begin{equation} \label{eqn:Hspin}
    \Hspin
        = \hbar \omegaS \Fz.
\end{equation}

\begin{figure}[t]
    \begin{center}
    \includegraphics[scale=1]{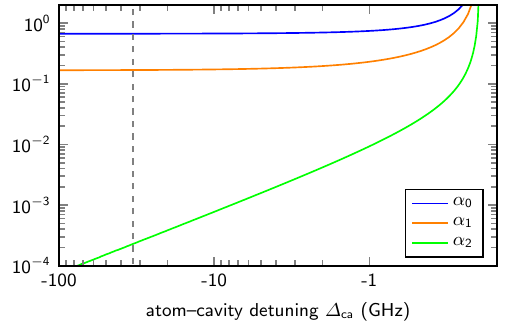}
    \end{center}

    \caption{ \label{fig:coupling-coefficients}
        Coupling coefficients describing the interaction given by \eqn{Hint-all-terms}.
        At $\DeltaCA / 2 \pi = \SI{-33.5}{\giga\hertz}$ (experimental value, gray dashed line), the tensor coupling $\alpha_2$ is negligible, and the scalar and vector coupling coefficients have reached their asymptotic values.
    }

\end{figure}

Generically, the interaction between the cavity and atom $i$ is described by
\begin{multline}
    \Hint^{(i)}
        = \hbar \sum_{g,e}
            g_{g;e}^+(\vec r_i) \,\hat c^\dag_+ \,\hat \sigma_{e;g}^{(i)} \,\delta_{m+1, m'} \\ {} +
            g_{g;e}^-(\vec r_i) \,\hat c^\dag_- \,\hat \sigma_{e;g}^{(i)} \,\delta_{m-1, m'} +
            \text{h.c.}
\end{multline}
Here, the summation runs over all possible transitions from the ground-state
manifold $g \equiv | f = 2, m \rangle$ to the excited states $e \equiv | f' = 3, m' \rangle$, with polarization-dependent coupling strengths $g_{g;e}^\pm$.
When the cavity--atom detuning $\DeltaCA$ is large compared to the hyperfine splittings $\varDelta_{f'}$ in the excited ($f' = 3$) manifold that is being addressed, the excited states can be eliminated.
This approximation results in a spin-dependent dispersive interaction Hamiltonian, describing dynamics within the ground-state manifold:
\begin{multline} \label{eqn:Hint-all-terms}
    \Hint^{(i)}
        = \hbar \gc |U(\vec r_i)|^2 \bigg\{
                \alpha_0 \big( \hat n_+ + \hat n_- \big) +
                \alpha_1 \big( \hat n_+ - \hat n_- \big) \hat f_k^{(i)} \\ {} +
                \alpha_2 \bigg[
                    \big( \hat n_+ + \hat n_- \big) \Big( \hat f_k^{(i)} \Big)^2 -
                    \hat c_- \hat c_+ \Big( \hat f_+^{(i)} \Big)^2 -
                    \hat c_+ \hat c_- \Big( \hat f_-^{(i)} \Big)^2
                \bigg]
            \bigg\},
\end{multline}
where $|U(\vec r_i)|^2$ is the local relative intensity of the cavity pump mode, where $\hat c_\pm$ are the annihilation operators for left- and right-handed cavity modes, which are approximately degenerate in our system, and where $\hat f_k$ and $\hat f_\pm$ are the spin operators relative to a quantization axis along the cavity axis $\hat k$.
Here, the scalar, vector, and tensor interactions between the spin and the cavity field are described by coupling coefficients $(\alpha_0, \alpha_1, \alpha_2) \to (2/3, 1/6, 0)$ in the limit of large $\DeltaCA$ (\fig{coupling-coefficients}).
In our system, the atomic ensemble is primarily localized within a single antinode of the cavity pump field, which allows the local cavity field $U(\vec r_i)$ to be treated as approximately constant.
This leaves
\begin{equation}
    \Hint
        = \hbar \gc \Big\{
                \alpha_0 \Na \hat n +
                \alpha_1 \big( \hat n_+ - \hat n_- \big) \Fk
            \Big\},
\end{equation}
such that the total system Hamiltonian, in the limit $|\DeltaCA| \gg |\varDelta_{f'}|$, is given by
\begin{multline} \label{eqn:Htot}
    \hat H
        =   -\hbar \DeltaPC \hat n +
            \hbar \omegaS \Fz \\ {} +
            \hbar \gc \Big\{
                \alpha_0 \Na \hat n +
                \alpha_1 \big( \hat n_+ - \hat n_- \big) \Fk
            \Big\}.
\end{multline}
When the cavity is pumped with only right-handed ($\sigma_-$) light, this reduces to \eqn{hamiltonian-general} of the main text.
When the cavity is pumped with only left-handed light, the sign of the cavity--spin interaction flips, and with it the sign of the gain of the feedback system.

\newpage
\section{B. Nondestructive measurement of the collective atomic spin state}
\label{sec:appendix:spin-measurement}

When the externally applied magnetic field is parallel to the cavity axis ($\thetaB = \SI{0}{\degree}$), the system Hamiltonian (\eqn{hamiltonian-general} of the main text), corresponding to pumping the cavity with right-handed ($\sigma_-$) light commutes with the total spin energy $\Fz$ since this becomes equivalent to the projection $\Fk$ of the spins along the cavity axis:
\begin{equation} \label{eqn:hamiltonian-longitudinal-rhc}
        \hat H^- =
        - \hbar \bigg( \DeltaPC - \frac 2 3 \gc \Na + \frac 1 6 \gc \Fz \bigg) \,\hat c^\dag \hat c
        + \hbar \omegaS \Fz,
\end{equation}
where the superscript on $\hat H^-$ indicates that this Hamiltonian only considers the right-handed cavity mode.
If the dispersive shift $\DeltaN$ to the cavity resonance condition is measured by comparing the resonance frequencies with and without the presence of atoms, it will be given by
\begin{equation} \label{eqn:deltaNplus}
    \DeltaN^-
        = - \frac 2 3 \gc \Na + \frac 1 6 \gc \langle \Fz \rangle,
\end{equation}
where the superscript on $\DeltaN^-$ indicates that this is the dispersive shift to the right-handed cavity mode.
If the atom number $\Na$ were known exactly, this measurement would be sufficient to determine the collective spin energy $\Fz$ of the ensemble; however, variable atom loss between state preparation and readout make this impractical.

By pumping the cavity with left-handed light, different information can be acquired.
Considering the case of an external field parallel to the cavity axis, the Hamiltonian can be derived which describes the left-handed ($\sigma_+$) cavity mode:
\begin{equation} \label{eqn:hamiltonian-longitudinal-lhc}
        \hat H^+ =
        - \hbar \bigg( \DeltaPC - \frac 2 3 \gc \Na - \frac 1 6 \gc \Fz \bigg) \,\hat c^\dag \hat c
        + \hbar \omegaS \Fz,
\end{equation}
which corresponds to a dispersive shift
\begin{equation} \label{eqn:deltaNminus}
    \DeltaN^+
        = - \frac 2 3 \gc \Na - \frac 1 6 \gc \langle \Fz \rangle.
\end{equation}

\begin{figure}[!bt]
    \begin{center}
    \includegraphics[scale=1]{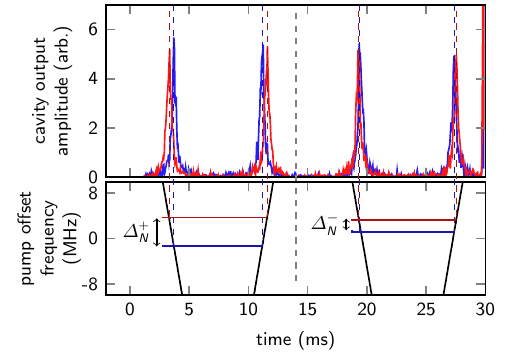}
    \end{center}

    \caption{ \label{fig:final-sweeps}
        An offset acousto-optic modulator (AOM) is used to sweep the frequency of the pump light back and forth across cavity resonance.
        This is repeated four times: twice with the atomic ensemble present in the cavity, and twice after the atoms have been expelled.
        For the first of each pair of sweeps, right-handed ($\sigma_-$) pump light is used; between sweeps, a pair of LCVRs is used to switch the polarization of the pump light to left-handed ($\sigma_+$) (grey dashed line).
        Comparing the resonance conditions with and without atoms present in the cavity offers a direct measure of $\DeltaN^\pm$, which can be used to calculate the final collective spin $\Fz$ and atom number $\Na$ of the ensemble.
    }

\end{figure}

Using a pair of liquid crystal variable retarders (LCVRs) at the input and the output of the cavity, the polarization of the light pumping the cavity can be switched rapidly between left- and right-handed without otherwise affecting the detection chain.
By measuring the resonance frequencies of each of the polarizations with the atomic ensemble present in the cavity, and then repeating both measurements with the atoms absent, the total atom number and collective spin can be recovered (\fig{final-sweeps}):
\begin{align}
    \Na                     & = -\frac 3 4 \frac{\DeltaN^+ + \DeltaN^-}{\gc}; \\
    \langle \Fz \rangle     & = -3 \frac{\DeltaN^+ - \DeltaN^-}{\gc}.
\end{align}
The same effect can be achieved by changing the orientation of the magnetic field to $\thetaB = \SI{180}{\degree}$ between the first and second measurements of $\DeltaN$, such that $\Fk = -\Fz$, but this takes too long to be practical due to the self-inductance of the coils used to generate the field.
The effect can also be achieved by using an external rf field to drive a $\pi$-pulse on the collective spin, taking $\Fz \to -\Fz$ between measurements; this approach has been successfully used in the past, but its dependence on the calibration of the rf drive makes it less appealing than switching the polarization of the pump light.

\newpage
\section{C. Quantum model of a collective spin coupled to an optical cavity}
\label{sec:appendix:model}

The damping rate $\dampingRate$ of the autonomous feedback system can be derived by considering how the system evolves in time.
Considering the Hamiltonian \eqn{hamiltonian-feedback} of the main text, and including terms accounting for pumping into and (non-Hermitian) leakage out of the cavity mode, the cavity field evolves according to
\begin{equation} \label{eqn:field-evolution}
    \frac{\dd}{\dd t} \hat c
        = \ii \big( \DeltaSlow + \gs \Fk \big) \hat c
            - \kappa \hat c
            + \kappa \eta,
\end{equation}
where $\eta$ is the coherent-state amplitude of the field pumping the cavity.
Here, the field operator $\hat c$ corresponds to the cavity field at frequency $\omegaP$.

Without any coupling to the cavity, $\gs = 0$, and \eqn{field-evolution} can be solved directly, giving
\begin{equation}
    \hat c_0
        = \eta \frac{\kappa}{\kappa - \ii \DeltaSlow}.
\end{equation}
If the effects of the coupling between the cavity and the collective spin are small, the perturbation to the field can be approximated by
\begin{equation} \label{eqn:field-rotating-frame}
    \hat c(t)
        = \hat c_0 + \hat c'(t).
\end{equation}

The collective spin, meanwhile, evolves according to
\begin{equation} \label{eqn:spin-evolution}
\begin{aligned}
    \frac{\dd}{\dd t} \Fx
        & = -\omegaS \hat F_y
            + \gs \hat c^\dag \hat c \Fy \cos \thetaB; \\
    \frac{\dd}{\dd t} \hat F_y
        & = \omegaS \Fx
            + \gs \hat c^\dag \hat c \big( \Fz \sin \thetaB - \Fx \cos \thetaB \big); \\
    \frac{\dd}{\dd t} \Fz
        & = - \gs \hat c^\dag \hat c \Fy \sin \thetaB.
\end{aligned}
\end{equation}
For $\omegaS \gg |\gs|$, as in our experiment, the transverse components admit solutions $\hat F_y \propto F_\perp \sin \omegaS t$.
For spins precessing at frequency $\omegaS$ at polar angle $\chi$ and azimuthal angle $\phi = \omegaS t$, the projection of the spin along the cavity axis looks like $\Fk = F_\perp \sin \thetaB \cos \phi + \Fz \cos \thetaB$, where $F_\perp \equiv F \sin \chi$ and $F_z \equiv F \cos \chi$.
Substituting this into \eqn{field-evolution} gives
\begin{multline} \label{eqn:field-equation-of-motion}
    \frac{\dd}{\dd t} \hat c'
        = \ii \big(
                \DeltaSlow
                + \gs F_\perp \sin \thetaB \cos \omegaS t
                + \gs \Fz \cos \thetaB
            \big) \big( \hat c_0 + \hat c' \big) \\
            - \kappa \big( \hat c_0 + \hat c' \big)
            + \kappa \eta.
\end{multline}
To lowest order, it seems reasonable to expect a solution that looks like effective cavity drives at frequencies $\pm \omegaS$ due to the modulation of the bare pump field $\hat c_0$ by the precessing spins.
This leads to the ansatz
\begin{equation}
    \hat c'(t)
        = \hat c_+ e^{\ii \omegaS t} + \hat c_- e^{-\ii \omegaS t}.
\end{equation}
Plugging this into \eqn{field-equation-of-motion} and ignoring quickly rotating terms $\sim e^{\pm 2 \ii \omegaS t}$ as well as terms of order $[(\gs / 2 \kappa) F_\perp \sin \thetaB]^2$ gives
\begin{align}
    \hat c_0 \label{eqn:carrier-amplitude}
        & = \eta \mathcal L(\DeltaSlow + \gs F_z \cos \thetaB); \\
    \hat c_+ \label{eqn:ansatz-a}
        & = \frac \ii 2 \frac \gs \kappa  F_\perp \sin \thetaB
            \,\mathcal L(\DeltaSlow + \gs F_z \cos \thetaB + \omegaS)
            \,\hat c_0; \\
    \hat c_- \label{eqn:ansatz-b}
        & = \frac \ii 2 \frac \gs \kappa  F_\perp \sin \thetaB
            \,\mathcal L (\DeltaSlow + \gs F_z \cos \thetaB - \omegaS)
            \,\hat c_0.
\end{align}
Here, $\mathcal L(\varDelta) \equiv \kappa / (\kappa - \ii \varDelta)$ refers to a Lorentzian line at center frequency $\varDelta$ with width $\kappa$.

It is desirable to find the effect of the cavity field on the spin energy $F_z$.
This is given by \eqn{spin-evolution}, and depends on the instantaneous occupation number $\hat n \equiv \hat c^\dag \hat c$ of the cavity mode:
\begin{equation}
\begin{aligned}
    \hat n
        & = \hat n_0 +
            \big( \hat c^\dag_+ \,\hat c_0 + \hat c_0^\dag \,\hat c_- \big) \,e^{-\ii \omegaS t} +
            \big( \hat c^\dag_- \,\hat c_0 + \hat c_0^\dag \,\hat c_+ \big) \,e^{\ii \omegaS t} \\
        & = \hat n_0 -
            \frac \ii 2 \frac \gs \kappa F_\perp \sin \thetaB \hat n_0 \\
             & \hspace*{3em} {} \times \Big[
                \mathcal L(-\DeltaSlow - \gs F_z \cos \thetaB + \omegaS)
                    \,e^{-\ii \omegaS t} \\ & \hspace*{4em} {} -
                \mathcal L (\DeltaSlow + \gs F_z \cos \thetaB + \omegaS)
                    \,e^{-\ii \omegaS t} \\ & \hspace*{4em} {} +
                \mathcal L (-\DeltaSlow - \gs F_z \cos \thetaB - \omegaS)
                    \,e^{\ii \omegaS t} \\ & \hspace*{4em} {} -
                \mathcal L(\DeltaSlow + \gs F_z \cos \thetaB - \omegaS)
                    \,e^{\ii \omegaS t}
            \Big]
\end{aligned}
\end{equation}
Again, terms of order $(\omegaS / \kappa)^2$, corresponding to second-order sidebands, have been ignored.
Using the cycle-averages $\overline{e^{\pm \ii \phi} \sin \phi} = \pm \ii / 2$, this gives the mean change in energy of the collective spin to be
\begin{equation} \label{eqn:change-in-Fz}
\begin{aligned}
    \overline{\frac{\dd}{\dd t} \langle \hat F_z \rangle}
        & = -\frac 1 2 F_\perp^2 \sin^2 \thetaB
            \frac{\gs^2}{\kappa} \big\langle \hat c_0^\dag \hat c_0 \big\rangle \\ & \hspace{1em} {} \times
            \Big[
                \Real \mathcal L(\DeltaSlow + \gs F_z \cos \thetaB + \omegaS) \\ & \hspace*{2em} {} -
                \Real \mathcal L(\DeltaSlow + \gs F_z \cos \thetaB - \omegaS)
            \Big],
\end{aligned}
\end{equation}
where the overline indicates time averaging over a Larmor precession cycle.
As expected, for $\DeltaSlow < - \gs F_z \cos \thetaB$, the first Lorentzian term is larger, and $F_z$ decreases; conversely, $F_z$ increases for $\DeltaSlow > - \gs F_z \cos \thetaB$.

\begin{figure}
    \begin{center}
    \includegraphics[scale=1]{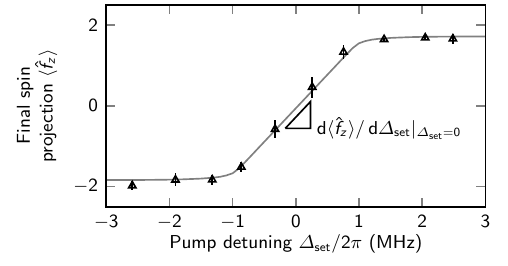}
    \end{center}

    \caption{ \label{fig:s-curve-fits}
        The coupled differential equations \eqn{gain-propagation} are propagated forward in time, and the results are used as a model function to fit to the measured values of $F_z$ (black triangles) for varying setpoints $\DeltaSlow$.
        The output of the fit (gray line) can be used to calculate the sensitivity of the autonomous feedback system to input noise.
    }

\end{figure}

The damping rate of the feedback system can be calculated as the exponential rate at which it approaches resonance in the limit $\gs F_z \cos \thetaB \to - \DeltaSlow$.
In the unresolved sideband regime $\omegaS \ll \kappa$, the asymmetry between the sidebands reduces to
\begin{multline}
    \Real \mathcal L(\DeltaSlow + \gs F_z \cos \thetaB + \omegaS) \\
    {} - \Real \mathcal L(\DeltaSlow + \gs F_z \cos \thetaB - \omegaS) \\
        {} \approx -\frac{4 \omegaS (\DeltaSlow + \gs F_z \cos \thetaB)}{\kappa^2}.
\end{multline}
Using this approximation along with \eqn{change-in-Fz} and \eqn{carrier-amplitude}, and noting that $\eta^2 = \nmax$ corresponds to the mean on-resonance photon occupation of the cavity, the damping rate of the system looks like
\begin{equation} \label{eqn:gain-final}
\begin{aligned}
    \dampingRate
        & = - \frac{\gs \cos \thetaB}{\DeltaSlow + \gs F_z \cos \thetaB}
            \overline{\frac{\dd}{\dd t} \langle \hat F_z \rangle} \\
        & = - 2 F_\perp^2 \sin^2 \thetaB \cos \thetaB
            \frac{\gs^3}{\kappa^3} \omegaS \nbar,
\end{aligned}
\end{equation}
where $\nbar = \nmax \,\kappa^2 / (\kappa^2 + [\DeltaSlow + \gs F_z \cos \thetaB]^2)$ is the true cavity-filtered photon occupation of the cavity.
For the parameters used in our experiment, this amounts to a damping rate of $\dampingRate = \SI{1600}{\per\second}$.
Notably, however, these parameters do not fall well within the low-modulation regime used to approximate \eqn{carrier-amplitude}.

The inclusion of higher-order terms $[(\gs / 2 \kappa) F_\perp \sin \thetaB]^2$ has the effect of reducing the carrier amplitude found in \eqn{carrier-amplitude}.
In particular, the full expression for the amplitude looks like
\begin{multline}
    \hat c_0
        = \eta \tilde{\mathcal L}(0) \\
            {} \times \Big[
                1 +
                \Big( \frac{\gs F_\perp \sin \thetaB}{2 \kappa}  \Big)^2
                \tilde{\mathcal L}(0)
                \big[ \tilde{\mathcal L}(\omegaS) + \tilde{\mathcal L}(-\omegaS) \big]
            \Big]^{-1},
\end{multline}
where $\tilde{\mathcal L}(\nu) \equiv \mathcal L(\DeltaSlow + \gs F_z \cos \thetaB + \nu)$ has been written for brevity.
For the parameters used in our experiment, this amounts to a correction factor of $0.7$, resulting in a correction factor of $0.5$ to $\nbar$ and to the final damping rate: $\dampingRate = \SI{790}{\per\second}$.
The solutions given by \eqn{ansatz-a} and \eqn{ansatz-b} still ignore quickly rotating terms corresponding to higher-order sidebands; however, these effects are confirmed experimentally to be small.

In order to simulate the dynamics of the system, $F_z \equiv \langle \hat F_z \rangle$ can be treated as a c-number and \eqn{gain-final} can be used to propagate $F_z$ forward in time.
In the absence of any dephasing, this treatment can be made complete by requiring that the total spin is conserved, $F_z^2 + F_\perp^2 = 4 \Na^2$.
Dephasing can be included heuristically by decreasing $F_\perp$ over time.
In practice, this decrease can take many functional forms, but a simple exponential decay captures much of the system dynamics.
Simulating the feedback process, then, amounts to propagating two coupled differential equations:
\begin{equation} \label{eqn:gain-propagation}
\begin{aligned}
    \frac{\dd}{\dd t} F_z
        & = \beta(F_\perp) \,F_z; \\
    \frac{\dd}{\dd t} F_\perp
        & = -\frac{\beta(F_\perp) \,F_z^2}{F_\perp} - \Gamma F_\perp.
\end{aligned}
\end{equation}
The resulting values of $F_z$ can be used as a model function for least-squares fitting, where $\Gamma$, as well as an overall offset to $F_z$ which accounts for systematic measurement errors, are allowed to vary (\fig{s-curve-fits}).
These fits are used to extract the sensitivities reported in \fig{dc-pulling}d.

\end{document}